\documentclass[twocolumn,aps,showpacs]{revtex4}
\usepackage{amssymb}
\usepackage{epsf}
\usepackage{amsmath}
\usepackage{graphicx}

\def\lsim{\lower -0.3ex \hbox{$<$} \kern -0.75em \lower 0.7ex \hbox{$\sim$}}
\def\gsim{\lower -0.3ex \hbox{$>$} \kern -0.75em \lower 0.7ex \hbox{$\sim$}}
\def\Journal #1,#2,#3,#4#5#6#7{#1 {\bf #2}, #3 (#4#5#6#7)}

\def\Vec#1{{\bf #1}}

\def\vare{\varepsilon}

\begin{document}

\title{
Multilayer graphenes with mixed stacking structure
\\
--- interplay of Bernal and rhombohedral stacking
}
\author{Mikito Koshino$^{1}$ and Edward McCann$^{2}$}
\affiliation{
$^{1}$Department of Physics, Tohoku University, Sendai, 980-8578, Japan\\
$^{2}$Department of Physics, Lancaster University, Lancaster, LA1
4YB, UK}

\begin{abstract}
We study the electronic structure of multilayer graphenes
with a mixture of Bernal and rhombohedral stacking
and propose a general scheme to understand the electronic band structure
of an arbitrary configuration.
The system can be viewed as a series of
finite Bernal graphite sections
connected by stacking faults.
We find that the low-energy eigenstates are
mostly localized in each Bernal section, and,
thus, the whole spectrum is well approximated by a collection of
the spectra of independent sections.
The energy spectrum is categorized
into linear, quadratic and cubic bands
corresponding to specific eigenstates of Bernal sections.
The ensemble-averaged spectrum exhibits
a number of characteristic discrete structures
originating from finite Bernal sections or their combinations
likely to appear in a random configuration.
In the low-energy region, in particular,
the spectrum is dominated by frequently-appearing
linear bands and quadratic bands
with special band velocities or curvatures.
In the higher energy region,
band edges frequently appear at some particular energies,
giving optical absorption edges at corresponding characteristic photon frequencies.
\end{abstract}


\maketitle

\section{Introduction}

\begin{figure}
\centerline{\epsfxsize=0.95\hsize \epsffile{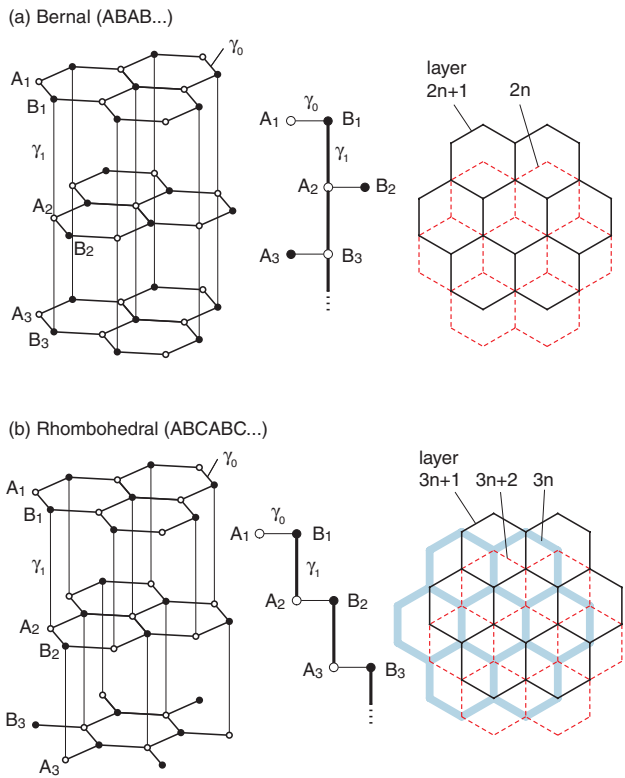}}
\caption{
Lattice structures of (a) Bernal graphite and
(b) rhombohedral graphite.
In each panel, the right figure is a top-view,
the middle is a schematic diagram of the
lattice structure.
}
\label{fig_aba_abc}
\end{figure}

Graphene multilayers exhibit
a wide variety of stacking arrangements allowed by
weak van der Waals interlayer coupling,
offering various types of quasiparticles
in the low-energy electronic spectrum.
In three-dimensional bulk graphite, there are two distinct
crystal configurations called Bernal ($ABAB\cdots$)
\cite{Wallace_1947a,McClure_1956a,Slonczewski_and_Weiss_1958a,McClure_1960a,Dres65,Dresselhaus_and_Dresselhaus_2002a},
and rhombohedral
($ABCABC\cdots$) stacking  \cite{Lipson_and_Stokes_1942a,Haering_1958a,Mcclure_1969a}
as illustrated in Fig.\ \ref{fig_aba_abc}.
A sequence such as $ABC\cdots$ represents the lattice point
on every layer along a perpendicular axis,
where $A$ and $B$ are inequivalent sublattices
of hexagonal lattice, and $C$ is the center of the hexagon.
Recently, several experimental techniques,
such as optical absorption
\cite{Nair_et_al_2008a,Kuzmenko_et_al_2008a,Mak_et_al_2010a,Mak_et_al_2010b},
Raman spectroscopy \cite{Ferrari_et_al_2006a,Lui_et_al_2011a}
and transmission electron microscopy \cite{Ping_and_Fuhrer_2012},
have been applied to identify the number of layers and the stacking order of
graphene multilayers. Few-layer graphene samples exfoliated from bulk graphite
usually exhibit Bernal structure which is supposed to be the most stable,
but often also display rhombohedral structure in part \cite{Lui_et_al_2011a,Ping_and_Fuhrer_2012}.

The electronic band structure of graphene multilayer
depends sensitively on its stacking structure \cite{Craciun_et_al_2009a,Kumar_et_al_2011a,Lui_et_al_2011a,Bao_et_al_2011a,Zhang_et_al_2011a,Taychatanapat_et_al_2011a,Khodkov_et_al_2012a,Henriksen_we_al_2012a,Ping_and_Fuhrer_2012}.
In Bernal stacked multilayer,
the spectrum consists of quadratic bands analogous to
bilayer graphene and a single linear band like monolayer
\cite{Guinea_et_al_2006a,Latil_and_Henrard_2006a,Aoki_and_Amawashi_2007a,Partoens_and_Peeters_2006a,Koshino_and_Ando_2007b,Koshino_and_Ando_2008a}.
In contrast, a rhombohedral-stacked multilayer has
a totally different spectrum with a pair of flat low-energy bands
which disperse as $p^N$ with momentum $p$ and the number of layers $N$
\cite{Guinea_et_al_2006a,Lu_et_al_2007a,Manes_et_al_2007a,Min_and_MacDonald_2008a,Koshino_and_McCann_2009b}.

\begin{figure}
\centerline{\epsfxsize=0.95\hsize \epsffile{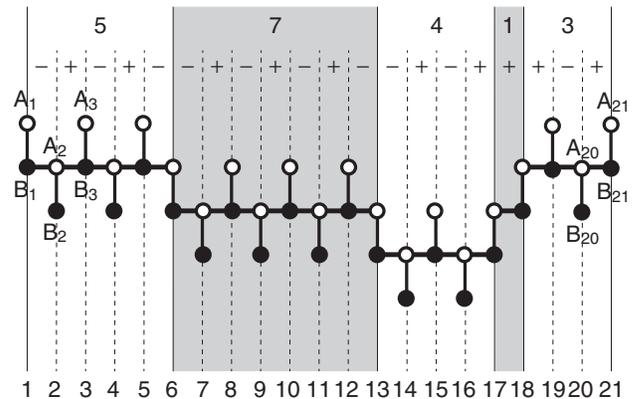}}
\caption{
Example of Bernal-rhombohedral mixed graphene multilayers
expressed by (5,7,4,1,3).
Intralayer coupling by parameter $\gamma_0$ is shown as a vertical solid
line, interlayer coupling by $\gamma_1$ is shown as a horizontal solid line.
}
\label{fig_atom}
\end{figure}

In general, graphitic structures are expected to take a
Bernal-rhombohedral mixed form as illustrated in Fig.\ \ref{fig_atom}.
The energy spectrum of mixed multilayer was
studied for some specific few-layer cases
\cite{Latil_and_Henrard_2006a,Aoki_and_Amawashi_2007a,Min_and_MacDonald_2008a},
but general rules predicting the electronic
properties of arbitrary structure are not well known.
A single rhombohedral stacking fault
appearing in Bernal graphite was studied theoretically \cite{Arovas_and_Guinea_2008a},
and it supports cubic bands
associated with the localized states bound to the stacking fault.
It was also shown \cite{Min_and_MacDonald_2008a} that the low-energy spectrum
of Bernal-rhombohedral mixed multilayer 
consists of energy bands which disperse as $p^J$,  
and the sum of $J$ coincides with the
number of layers.
This predicts the number of the bands belonging to
each $J$, but to obtain the quantitative
dispersions and wavefunctions
one would need to actually calculate eigen energies
for every single configuration.

In this paper, we study the electronic structures of
general Bernal-rhombohedral mixed graphene multilayers.
We begin by proposing a general scheme in which to understand the band property
of any given configuration
without resorting to diagonalizing the full Hamiltonian.
We view the system as a series of finite-layered
Bernal sections connected by the rhombohedral-type stacking fault,
as depicted in Fig.\ \ref{fig_fragment}(a),
and treat the coupling between neighboring sections as a perturbation.
We find that the eigenstates near the Dirac point are
mostly localized in each Bernal section,
and the states are approximated well by those of
incomplete Bernal graphites as illustrated in Fig.\ \ref{fig_fragment}(b)
and (c).
The energy spectrum is then categorized
into linear, quadratic and cubic (or higher order) bands
within the basis of incomplete Bernal graphite sections.
We also specify several limited situations
where the states of neighboring Bernal sections are strongly hybridized.

To model realistic experimental systems with many layers and, perhaps,
a number of stacking faults, we develop a statistical approach.
To study the electronic properties averaged
over different stacking configurations, we analyze the statistics of
the velocity of the linear bands and the
effective mass of the quadratic bands which dominate
the low-energy spectrum.
We find that there are some particular frequently-appearing
values in the velocity/mass distribution,
corresponding to finite-layered Bernal sections
and their particular combinations.
We also compute the averaged optical absorption spectrum
and find that the absorption edges emerge at particular frequencies,
corresponding to frequently-appearing band structures.

The paper is organized as follows.
We formulate the Hamiltonian of mixed multilayer graphene
in Sec.\ II.
We describe the eigenstates and the energy spectrum
of isolated incomplete Bernal graphite section,
and study the inter-section mixing effect in Sec.\ III.
We present the ensemble-averaged distribution of low-energy
band velocity and effective mass, and
the optical absorption in Sec. IV.

\section{Formulation and effective-mass Hamiltonian}

We consider an $N_{\rm tot}$-layer graphene system
which has a unit cell containing $A_j$ and $B_j$ sublattices
on the $j$-th layer.
Coupling between the $j$-th and $j+1$-th layers is
described as either $AB$ or $BA$ stacking,
where $AB$ stacking is defined as the arrangement
in which sites $A_j$ and $B_{j+1}$ are connected by
the vertical interlayer coupling, whereas  $BA$ stacking as the one
in which $B_j$ and $A_{j+1}$ are connected,
as illustrated Fig.\ \ref{fig_atom}.
The entire system is specified by a set of indices
$\sigma_j = \pm$ for $j=1,2,\cdots,N_{\rm tot}-1$
describing the interlayer connection between $j$-th and $j+1$-th layers,
where $+$ and $-$ represent $AB$ and $BA$ stacking, respectively.
Bernal-stacked graphene is then expressed as an alternating sequence like
$(+,-,+,-,\cdots)$, while rhombohedral-stacked graphene is
$(+,+,+,\cdots)$ or $(-,-,-,\cdots)$.

In a general configuration,
an alternative sequence $(\cdots,+,-,+,-,\cdots)$
is regarded as a section of continuous Bernal structure,
and a position at which the same sign
 $(\cdots,+,+,\cdots)$ or $(\cdots,-,-,\cdots)$
occurs consecutively is regarded as a rhombohedral-type stacking fault
separating different Bernal section.
The sequence $\{\sigma_i\}$ is alternatively
expressed as a set of integers
\begin{equation}
(N_1,N_2,N_3,\cdots,N_M),
\end{equation}
where $N_i$ is the length of the $i$-th Bernal section,
and the stacking fault exists between $N_i$ and $N_{i+1}$.
For example, the sequence $(+,-,+,-,-,+,-,-,+)$
is written as $(4,3,2)$.
$M$ is the number of separated sections in the whole system.
The total number of layers in the system is given by
\begin{equation}
 N_{\rm tot} = 1+  \sum_{i=1}^M N_i.
\end{equation}
A pure Bernal-stacked multilayer graphene is represented
by a single number $(N_{\rm tot}-1)$, and a pure
rhombohedral multilayer is by $(1,1,1,\cdots)$.

To describe the electronic properties,
we use an effective-mass model
\cite{McClure_1956a,Slonczewski_and_Weiss_1958a,DiVincenzo_and_Mele_1984a,Semenoff_1984a,Shon_and_Ando_1998a,Ando_2005a}
with the Slonczewski-Weiss-McClure parameterization of graphite \cite{Dresselhaus_and_Dresselhaus_2002a}.
As the simplest approximation, we include parameter $\gamma_0$
describing the nearest neighbor coupling within each layer,
and $\gamma_1$ for the coupling of the interlayer vertical bonds.
The band parameters were experimentally estimated
in the bulk ABA graphite, for example \cite{Dresselhaus_and_Dresselhaus_2002a} 
as  $\gamma_0=3.16$ eV and $\gamma_1 = 0.39$ eV.
The low energy spectrum is given by states
in the vicinity of the $K_\xi$ point at the corner of the Brillouin zone,
where $\xi = \pm 1$ is the valley index.
If $|A_j\rangle$ and $|B_j\rangle$ are Bloch functions at the $K_{\xi}$
point, corresponding to the $A$ and $B$ sublattices of layer $j$,
respectively, then, in the basis of
$|A_1\rangle,|B_1\rangle,|A_2\rangle,|B_2\rangle$, $\cdots$,
the Hamiltonian in the vicinity of the $K_{\xi}$ valley is
\begin{eqnarray}
H =
\begin{pmatrix}
 H_0 & V_1  \\
 V_1^\dagger & H_0 & V_2 \\
  & V_2^\dagger & H_0 & V_3  \\
  &  & V_3^\dagger & H_0 & V_4  \\
  & &  & \ddots & \ddots & \ddots
\end{pmatrix},
\label{eq_H}
\end{eqnarray}
with
\begin{eqnarray}
&& H_0 =
\begin{pmatrix}
 0 & v \pi^\dagger \\ v \pi & 0
\end{pmatrix},
\label{Hdef}
\\
&& V_j =
\left\{
\begin{array}{l}
\begin{pmatrix}
 0 & 0 \\ \gamma_1 & 0
\end{pmatrix} \quad (\sigma_j=-),
 \\
\begin{pmatrix}
 0 & \gamma_1 \\ 0 & 0
\end{pmatrix} \quad (\sigma_j=+).
\end{array}
\right.
\label{Vdef}
\end{eqnarray}
Here, the in-plane momentum operator is $\pi = \xi \hat{p_x} + i \hat{p_y}$,
and $\hat{\Vec{p}} = \left(\hat{p}_x , \hat{p}_y \right) = -i\hbar \nabla$.
The diagonal blocks, Eq.~(\ref{Hdef}), describe nearest-neighbor intralayer
hopping, and $V$, Eq.~(\ref{Vdef}), describes nearest-neighbor layer
hopping. $v$ is the band velocity of monolayer graphene
given by $v = \sqrt{3} a \gamma_0/2\hbar$,
where $a \approx 0.246$ nm is the lattice constant of honeycomb lattice.
We neglect other hopping parameters in the following arguments
for simplicity. We expect that the neglected parameters
introduce relatively small corrections such as trigonal warping
and electron-hole asymmetry, as in multilayer graphenes
with pure Bernal
\cite{Dresselhaus_and_Dresselhaus_2002a,McCann_and_Falko_2006a,Koshino_and_Ando_2007b, Partoens_and_Peeters_2006a}
and rhombohedral stacking \cite{Mcclure_1969a,Koshino_and_McCann_2009b}.

\section{Electronic structure and band classification}

Unlike a pure Bernal multilayer,
the Hamiltonian of a general graphene stack
cannot be simply decomposed into smaller subsystems
by a unitary transformation.
At small momentum, however,
we can show that every eigenstate is almost
well localized in a single Bernal-graphite section between
rhombohedral stacking faults,
so that the system can be treated approximately
as a set of independent Bernal sections.

\begin{figure}
\centerline{\epsfxsize=0.85\hsize \epsffile{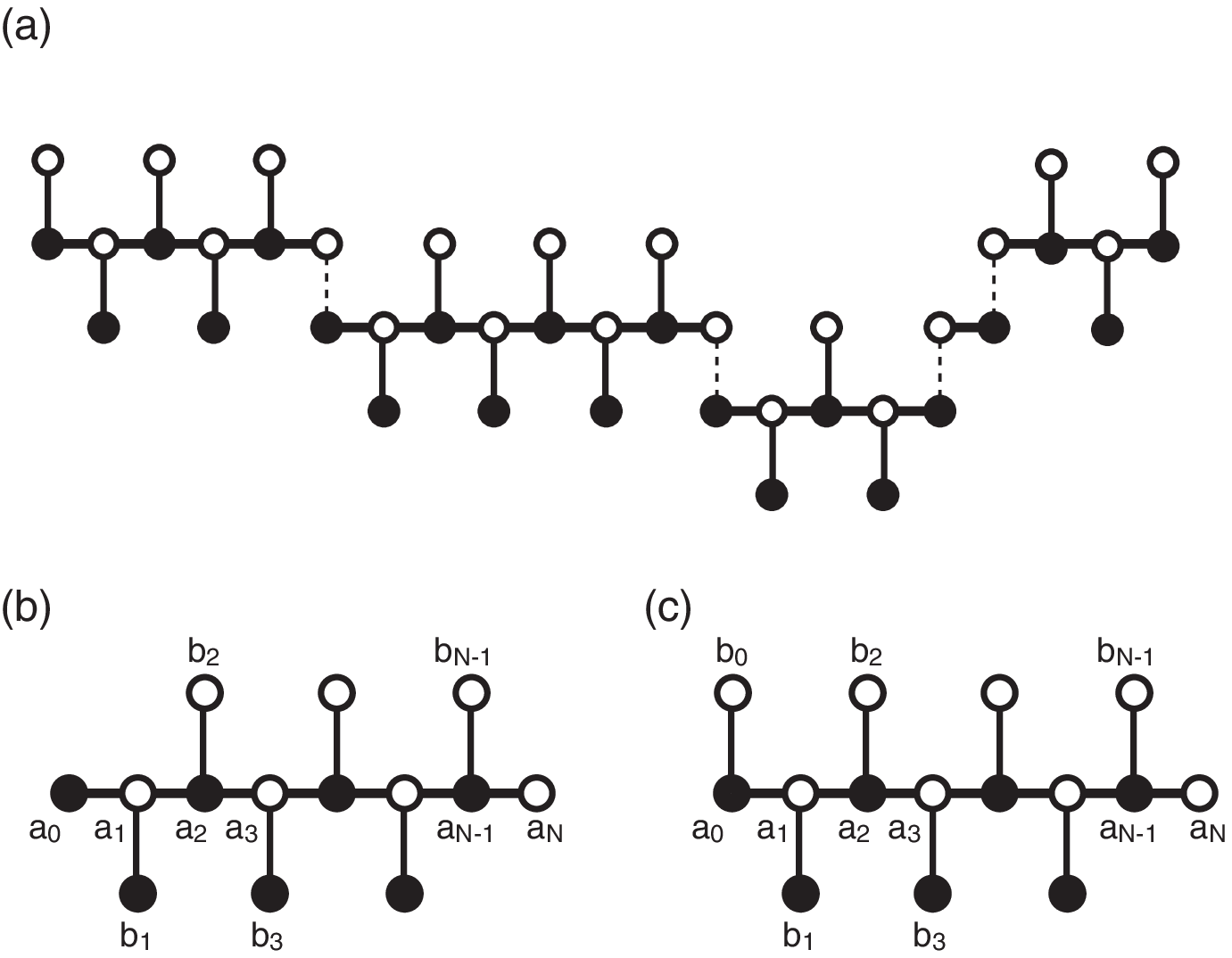}}
\caption{
(a) Decomposition of Bernal-rhombohedral mixed multilayers
into incomplete Bernal sections.
Dashed lines are connections between neighboring sections,
which are to be switched off in the decomposition.
(b) Decomposed incomplete Bernal section in the middle
and (c) at one end of the whole system.
}
\label{fig_fragment}
\end{figure}

In order to develop an analytical description, we begin
by considering an imaginary system
in which the {\em intralayer} hopping is switched off
only on the stacking-fault layers, as illustrated
in Fig.\ \ref{fig_fragment}(a).
The entire system is then broken into a set of
incomplete Bernal graphite sections
with two inequivalent sublattice sites per layer
but with one sublattice missing at the stacking fault layers,
as illustrated in Fig.\ \ref{fig_fragment}(b) and (c)
for a middle section and an end section, respectively.
In the following, we will show that at small momentum
$p  \ll \gamma_1/v$,
the spectrum is approximately that of a collection of
the energy bands of isolated incomplete Bernal graphites,
except for some specific occasions where
the coupling between different sections is significant.

\subsection{Spectrum of incomplete graphite}

We first consider the electronic structure of
an isolated incomplete Bernal graphite of the middle section type, as
in Fig.\ \ref{fig_fragment}(b).
For convenience, we divide all the atomic sites into two groups,
`chained' sites and `free' sites, where
a chained site refers to a site connected to neighboring layers
by $\gamma_1$, and a free site
to one which has only intralayer connection $\gamma_0$.
This incomplete Bernal middle section has $N+1$ layers that may be numbered
from $j=0$ to $N$. In this respect it is identical to Bernal-stacked
graphite, but, as compared to Bernal-stacked graphite, there is a missing
site at each end, and the missing sites are always free sites.
We rename the sites $A_j$ and $B_j$
with $a_j$ and $b_j$, so that
$a_j$ and $b_j $ represents the chained site
and the free site on the $j$-th layer, respectively.
If $a_0 = B_0$, for example, the relation is
\begin{eqnarray}
&&a_{j} =
\left\{
\begin{array}{c}
B_j \quad (j={\rm even})
\\
A_j \quad (j={\rm odd})
\end{array}
\right.,
\nonumber
\\
\quad
&&b_{j} =
\left\{
\begin{array}{c}
A_j \quad (j={\rm even})
\\
B_j \quad (j={\rm odd})
\end{array}
\right..
\end{eqnarray}

The Hamiltonian is obtained by
eliminating the missing sites ($b_0$ and $b_N$)
in complete $N+1$-layer Bernal graphite.
As the system includes $2N$ atoms,
there are $2N$ eigenfunctions at a momentum $\Vec{p}$,
each of which may be expressed as
\begin{eqnarray}
&&
\begin{pmatrix}
\Psi(a_j) \\
\Psi(b_j)
\end{pmatrix}
=
\begin{pmatrix}
f(a_j) \\
f(b_j)
\end{pmatrix}
e^{i \Vec{p}\cdot \Vec{x}},
\end{eqnarray}
where $\Vec{x} = (x,y)$ is the in-plane position and
$\Vec{p} = (p_x,p_y)$ is the in-plane momentum
measured from $K_\xi$.
In this simple model with $\gamma_0$ and $\gamma_1$,
the energy bands are always isotropic around $p=0$,
and the dispersion is a function of $p\equiv \sqrt{p_x^2+p_y^2}$.

\begin{figure*}
\centerline{\epsfysize=0.4\vsize \epsffile{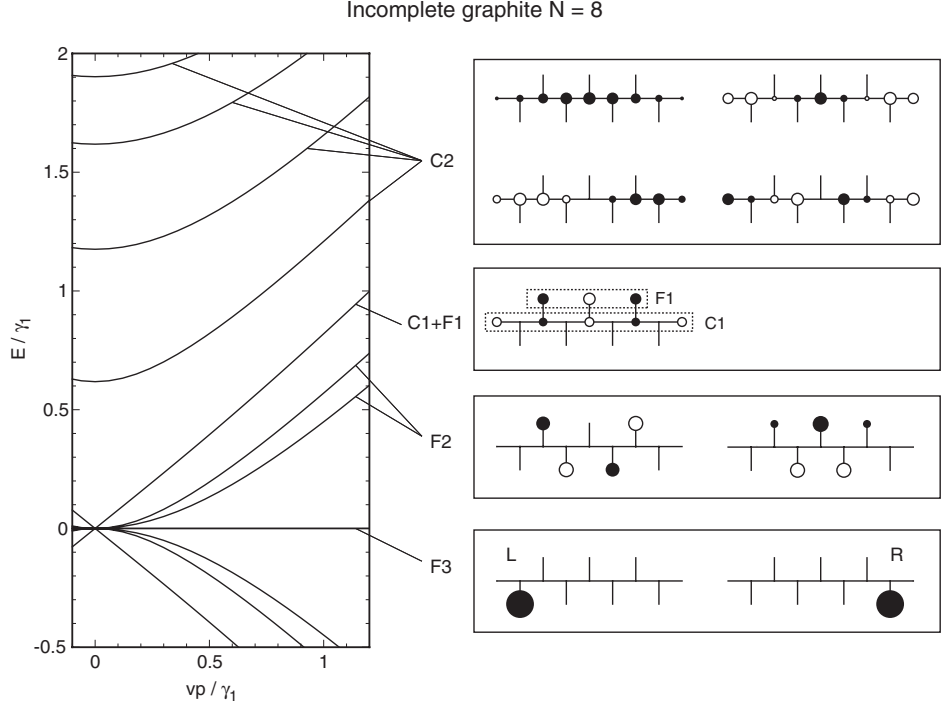}}
\caption{(Left) Energy bands of incomplete Bernal graphite with $N=8$.
(Right) Some of wavefunctions from different categories
at a small momentum along $p_x$ direction with $|p_x|\ll \gamma_1/v$.
White and black circles represent positive and negative
wave amplitudes, respectively, and the size is the absolute magnitude.
}
\label{fig_band_8}
\end{figure*}

As we describe in the following, it is possible to
classify the bases of the eigenfunctions into five categories:
\begin{eqnarray}
&& \mbox{C1: chained, linear}\nonumber\\
&& \mbox{C2: chained, quadratic}\nonumber\\
&& \mbox{F1: free, linear}\nonumber\\
&& \mbox{F2: free, quadratic}\nonumber\\
&& \mbox{F3: free, boundary-localized}.
\end{eqnarray}
C and F represent chained and free sites, respectively.
1 and 2 correspond to the linear and quadratic dispersion
in the band structure.
F3 gives a dispersion-less flat band in the isolated incomplete graphite,
but it forms to a cubic (or higher order)
band when the inter-section coupling is included.
As an example, we present in Fig.\ \ref{fig_band_8}
the band structure and
the wave functions of incomplete graphite $N=8$.

The bases of C1 and C2 are
plane waves on the chained sites,
which vanish at imaginary sites
$a_{-1}$ and $a_{N+1}$ outside the system.
Explictly, this is defined as
\begin{eqnarray}
&&
\begin{pmatrix}
f^{\rm C}_{q_l}(a_j) \\
f^{\rm C}_{q_l}(b_j)
\end{pmatrix}
=
\begin{pmatrix}
\sqrt{\frac{2}{N+2}}\sin q_l(j+1)\\
0
\end{pmatrix},
\label{eq_wave_C}
\end{eqnarray}
with quantized wave numbers
\begin{eqnarray}
&& q_l = \frac{l \pi}{N+2}, \quad l=1,2,\cdots,N+1.
\label{eq_q_C}
\end{eqnarray}
The wavenumber $q_l = \pi/2$, appearing only when $N$ is even,
is categorized in C1, and all the others in C2.

The bases of F1 and F2 are plane waves on the free sites,
but they have nodes at $b_1$ and $b_{N-1}$ inside the system. They are written as
\begin{eqnarray}
&&
\begin{pmatrix}
f^{\rm F}_{q'_l}(a_j) \\
f^{\rm F}_{q'_l}(b_j)
\end{pmatrix}
=
\begin{pmatrix}
0 \\
e^{i\theta_j} \sqrt{\frac{2}{N-2}}\sin q'_l (j-1)\\
\end{pmatrix},
\label{eq_wave_F}
\end{eqnarray}
with a different series of wave numbers
\begin{eqnarray}
&& q'_l = \frac{l \pi}{N-2}, \quad
l=1,2,\cdots,N-3,
\label{eq_q_F}
\end{eqnarray}
where $\theta_j$ is defined by $e^{i\theta_j} = (p_x\pm i p_y)/p$
when $b_j = B_j$ and $A_j$, respectively.
F1 or F2 states do not exist when $N\leq 3$.
The wavenumber $q'_l = \pi/2$ which appears
when $N$ is even (more than 4) is categorized in F1,
and all the others in F2.

The group F3 is comprised of two states localized at
$b_1$ or $b_{N-1}$,
\begin{eqnarray}
&&
\begin{pmatrix}
f^{\rm F3}_{\rm L}(a_j) \\
f^{\rm F3}_{\rm L}(b_j)
\end{pmatrix}
=
\begin{pmatrix}
0 \\
\delta_{j,1}
\end{pmatrix},
\nonumber\\
&&
\begin{pmatrix}
f^{\rm F3}_{\rm R}(a_j) \\
f^{\rm F3}_{\rm R}(b_j)
\end{pmatrix}
=
\begin{pmatrix}
0 \\
\delta_{j,N-1}
\end{pmatrix}.
\label{eq_wave_F3}
\end{eqnarray}
In the case $N=2$, there is a single F3 state
localized at the only free site.
When $N=1$, there are no F3 states.

At $p=0$, C1 states and all F states (F1, F2 and F3)
are the exact eigenstates at zero-energy.
Considering the perturbation in $p$ for these degenerate states,
the Hamiltonian are approximately
block diagonalized into blocks describing C1+F1, F2 and F3 states.
C1 and F1 are hybridized by a term linear in $p$,
to form a monolayer-like Dirac cone,
\begin{eqnarray}
 && \vare^{\rm C1+F1}_\pm (p)
\approx  \pm \sqrt{\frac{N-2}{N+2}} vp,
\label{eq_band_C1+F1}
\end{eqnarray}
but with a reduced band velocity compared to monolayer graphene
\cite{Min_and_MacDonald_2008a}.
Mixing with other states gives rise to a
correction to Eq.\ (\ref{eq_band_C1+F1}), but it is quadratic in $p$.
The C1+F1 band only appears when $N$ is an even number, 
because otherwise C1 or F1 state does not exist.
The case of $N=2$ is special, in that
F1 does not exist and C1 alone gives a flat band
at zero energy.

F2 states give low-energy quadratic bands,
\begin{eqnarray}
\vare^{\rm F2}_{q'_l} (p) \approx
- \frac{v^2p^2}{2\gamma_1 \cos q'_l},
\label{eq_band_F2}
\end{eqnarray}
while F3 remains at zero energy independently of $p$,
\begin{eqnarray}
\vare^{\rm F3}_{\rm L} (p) = \vare^{\rm F3}_{\rm R} (p) = 0 .
\label{eq_band_F3}
\end{eqnarray}
In Appendix \ref{sec_app1}, we show that F2 and
F3 states are actually the approximate eigenstates
with the eigen energies Eqs.\ (\ref{eq_band_F2}) and
(\ref{eq_band_F3}), respectively.
F2 bands of $q'_l$ and $\pi-q'_l$
form a pair of electron and hole bands with the same band mass
$m^* = \gamma_1|\cos q'_l|/v^2$,
analogous to the low-energy bands of bilayer graphene
with the mass $m^* = \gamma_1/(2v^2)$ \cite{McCann_and_Falko_2006a}.
Indeed, the $2\times 2$ effective Hamiltonian spanned by these two F2 states
is shown to be equivalent with the low-energy Hamiltonian
for bilayer graphene by an appropriate unitary transformation.

C2 states are the eigenstates on the chained sites at $p=0$,
with the non-zero eigen energy $2\gamma_1 \cos q_l$.
For $p\neq 0$, those states are hybridized with the free sites
giving rise to quadratic dispersion for $vp \ll 2\gamma_1 \cos q_l$,
\begin{eqnarray}
&& \vare^{\rm C2}_{q_l} (p) \approx
2\gamma_1 \cos q_l +
\frac{v^2p^2}{2\gamma_1 \cos q_l}
\left(
1-\frac{4}{N+2}\sin^2 q_l
\right).\nonumber\\
\label{eq_band_C2}
\end{eqnarray}
This band is electron-type and hole-type for
$0 < q_l < \pi/2$ and $\pi/2 < q_l < \pi$, respectively.

In a section located at the end of the whole system
as in Fig.\ \ref{fig_fragment}(c),
one of missing sites, $b_0$ or $b_N$ is restored,
giving a different energy spectrum.
When $b_0(b_N)$ exists, the plane waves of free sites
in Eq.\ (\ref{eq_wave_F}) have a node at $b_{-1}(b_{N+1})$
(out of the system) instead of $b_1(b_{N-1})$,
so that the quantized wavenumbers on F states,
Eq.\ (\ref{eq_q_F}), are changed to
\begin{eqnarray}
&& q'_l = \frac{l \pi}{N}, \quad
l=1,2,\cdots,N-1,
\label{eq_q_F_end}
\end{eqnarray}
The dispersion of the F2 band in Eq.\ (\ref{eq_band_F2})
changes according to the new set of $q'_l$.
There are no changes for the C state wavenumber in Eq.\ (\ref{eq_q_C}).
The linear band of C1+F1, Eq. (\ref{eq_band_C1+F1}),
is modified to
\begin{eqnarray}
 && \vare^{\rm C1+F1}_\pm (p)
\approx  \pm \sqrt{\frac{N}{N+2}} vp.
\end{eqnarray}
For F3 states, either (F3,L) or (F3,R),
whichever is closer to the end of the whole system, does not exist any more.


When both of $b_0$ and $b_{N}$ are restored,
the partial system becomes a complete $(N+1)$-layer Bernal graphite,
and $q_l$ and $q'_l$ become identical to Eq.\ (\ref{eq_q_C}).
Then C2 and F2 bands belonging to $q_l$ and $\pi-q_l$
compose a subsystem equivalent to bilayer graphene,
and the C1+F1 band forms a linear band with band velocity
equal to the monolayer's \cite{Guinea_et_al_2006a,Koshino_and_Ando_2007b}.

\begin{figure}
\centerline{\epsfxsize=0.85\hsize \epsffile{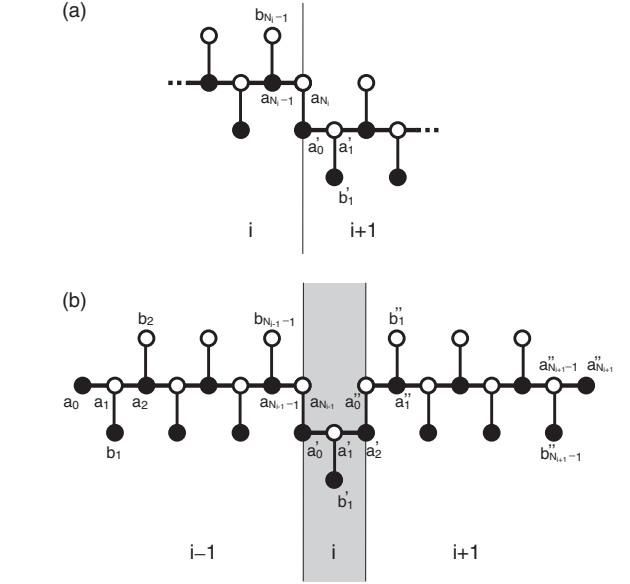}}
\caption{(a) Schematic of a stacking fault between the neighboring
Bernal sections.
(b) Structure of $(N_{i-1},2,N_{i-1})$.
}
\label{fig_connect}
\end{figure}

\subsection{Coupling beyond the stacking faults}

The wavefunctions of neighboring incomplete
graphite sections are generally hybridized at $p \neq 0$
via the bond between the $a_0$ site of one section
and $a_N$ of the other, as illustrated in Fig.\ \ref{fig_connect}(a).
Particularly strong coupling occurs between the C1+F1 band and the C2 band
because they have significant wave amplitudes at the end sites
$a_0$ and $a_N$,
\begin{eqnarray}
|f^{\rm C}_{q_l}(a_1)| = |f^{\rm C}_{q_l}(a_N)|
= \sqrt{\frac{2}{N+2}}\sin q_l.
\label{eq_endsite_amp}
\end{eqnarray}
The coupling is generally large
at $q_l = \pi/2$ (C1 state), and decreases
as $q_l$ approaches 0 or $\pi$.
The F2 and F3 bands, on the other hand,
are mainly localized on free sites
so that the mixing effect is generally small.
In the following we consider details of the inter-section
hybridization for each eigenstate class.

\subsubsection{C1+F1 band}

If the sequence of integers
$(N_1,N_2,N_3,\cdots,N_M)$ describing the
lengths $N_i$ of each Bernal section
contains any consecutive even numbers, then
the C1+F1 linear bands in those sections are mixed
together through the coupling among C1 states.
Let us assume $N_s,N_{s+1},\cdots,N_{s+p}$ are a series of consecutive
even numbers.
We arrange the bases as
$|s,{\rm C1}\rangle,|s,{\rm F1}\rangle,\cdots,|s+p,{\rm
C1}\rangle,|s+p,{\rm F1}\rangle$,
where $|i,{\rm C1}\rangle$ represents the C1 state of the section of $N_i$.
The Hamiltonian matrix is written for this basis as
\begin{eqnarray}
 H^{\rm C1+F1} =
\begin{pmatrix}
 h_s & u_{s} &  \\
u_{s}^\dagger & h_{s+1} & u_{s+1} \\
& u_{s+1}^\dagger & h_{s+2} & u_{s+2} \\
&  & \ddots & \ddots & \ddots \\
 & & & u_{s+p-1}^\dagger & h_{s+p}
\end{pmatrix},
\label{eq_C1+F1_mixed}
\end{eqnarray}
with
\begin{eqnarray}
&& h_i =
\sqrt{\frac{N_i-2(1-\delta_i)}{N_i+2}}
\begin{pmatrix}
0 & v \pi_i^\dagger \\
v \pi_i & 0
\end{pmatrix},
\\
&& u_i =
\sqrt{\frac{1}{(N_i+2)(N_{i+1}+2)}}
\begin{pmatrix}
v \pi_i^\dagger & 0\\
0 & 0
\end{pmatrix},
\end{eqnarray}
where $\pi_i = \pi (\pi^\dagger)$ when
$a_0$ of the section $s$ is $A(B)$ site,
and $\delta_i = 1$ when the section $N_i$
is located at the end of the entire system (i.e., $i=1$ or $M$),
and $\delta_i=0$ otherwise.

The Hamiltonian Eq.\ (\ref{eq_C1+F1_mixed}) immediately leads to
$p+1$ Dirac cones with generally different band velocities
from the original.
We found that the velocities are always equal to,
or smaller than the monolayer's band velocity $v$,
and that velocity equal to $v$ only appears
when $N_s, \cdots, N_{s+p}$ covers the whole system,
i.e., $s=1$ and $s+p = M$.

Fig.\ \ref{fig_band_6412} shows an example
of the band structure and the wavefunctions,
calculated for the graphene stack $(6,4,1,2)$.
Dashed curves plotted for $p<0$ and $E>0$ represent
the energy bands of the decomposed incomplete graphites, $N=6,4,1,$ and $2$.
We see that the linear bands of the neighboring sections $N=6$ and 4
are hybridized to give the 5th and 7th bands
which have different velocities from the original,
while the linear band from $N=2$ is kept almost intact (6th band).

\subsubsection{F2 band}

The coupling between F2 bands in the neighboring sections is
in proportional to $p^3$,
because the amplitude at $a_0$ and $a_N$ is of the order of $p$,
and the hopping between $a_0$ and $a_N$ is $vp$.
For small momenta $vp \ll \gamma_1$, the coupling effect
is negligible compared to the original F2 band energies $\propto p^2$,
so that the wavefunction of F2 states are well localized
in each single section.

An exceptionally strong inter-section coupling
occurs where a section of $N_i = 2$ exists in the middle of the system,
as illustrated in Fig.\ \ref{fig_connect}(b).
There C1 states of $N_i =2$
strongly hybridizes F2 states of the adjacent sections $i-1$ and $i+1$,
and also two F3 states  at $b_{N_{i-1}-1}$ and at $b''_{1}$
facing to the $N_i =2$.
The hybridized eigen energies
are given by Eq.\ (\ref{eq_band_F2}) with
reconstructed wavenumber $q_l'$, which are the solutions of
\begin{eqnarray}
&&
\sin \left[q'_l(N_{i-1}-1)\right]  \sin \left[q'_l(N_{i+1}-1)\right]
\cos^2 q'_l \nonumber\\
&&=
\cos \left[q'_l(N_{i-1}-1)\right]  \cos \left[q'_l(N_{i+1}-1)\right]
\sin^2 q'_l.
\label{eq_reconst1}
\end{eqnarray}
If the section $i\pm 1$ is the end section,
then $N_{i\pm 1}$ should be replaced with $N_{i\pm 1}+2$.
The detailed derivation of Eqs.\ (\ref{eq_reconst1})
is presented in the Appendix \ref{sec_app2}.

\subsubsection{F3 band}

Similarly to F2,
the coupling between F3 bands in neighboring sections is
proportional to $p^3$.
Since the F3 is originally a zero energy band in an isolated Bernal graphite,
the coupling of $O(p^3)$ directly becomes the band dispersion
when the inter-section coupling is introduced.
Let us we consider F3 states
$|i,{\rm F3,R}\rangle$ and $|i+1,{\rm F3,L}\rangle$
in the neighboring sections $i$ and $i+1$,
which are facing each other across the stacking fault.
We index the sites with $a_j,b_j$ and
$a'_j,b'_j$ for the section $i$ and $i+1$, respectively,
as illustrated in Fig.\ \ref{fig_connect}(a).
The state $|i,{\rm F3,R}\rangle$ is localized at
$b_{N_i-1}$, and $|i+1,{\rm F3,L}\rangle$ is at $b'_1$.
When $b_{N_i-1}$ is an $A$ site,
the effective Hamiltonian in the basis of
$|i,{\rm F3,R}\rangle$ and $|i+1,{\rm F3,L}\rangle$ becomes,
\begin{eqnarray}
&& \delta {\cal H}^{\rm (eff)}  =
\begin{pmatrix}
0 & (v \pi^\dagger)^3/\gamma_1^2 \\
(v \pi)^3/\gamma_1^2 & 0
\end{pmatrix}
\label{eq_H_F3},
\end{eqnarray}
leading to cubic dispersion of $\vare = \pm v^3p^3/\gamma_1^2$.
This actually agrees with the cubic band of the stacking-fault bound states
argued in Ref. \cite{Arovas_and_Guinea_2008a}.
When $b_{N_i-1}$ is a $B$-site,
$\pi$ and $\pi^\dagger$ are interchanged in Eq.\ (\ref{eq_H_F3}).

The Hamiltonian Eq.\ (\ref{eq_H_F3}) is quite similar to
that of low-energy states in rhombohedral trilayer graphene \cite{Guinea_et_al_2006a,Manes_et_al_2007a,Min_and_MacDonald_2008a,Koshino_and_McCann_2009b}.
If the stacking faults appear $s$ times consecutively,
i.e., $(\cdots,N_i,1,\cdots,1,N_{i+s},\cdots)$, then
the states $|i,{\rm F3,R}\rangle$ and $|i+s,{\rm F3,L}\rangle$
forms a Hamiltonian equivalent to rhombohedral $(s+2)$-layer,
giving $p^{s+2}$ dispersion.
In the multilayer (6,4,1,2), we have
a pair of $p^3$ bands (2nd band) at the boundary between $N=6$ and 4,
and $p^4$ bands (1st band) between $N=4$ and 2,
as shown in Fig.\ \ref{fig_band_6412}.

\subsubsection{C2 band}

\begin{figure*}
\centerline{\epsfysize=0.45\vsize \epsffile{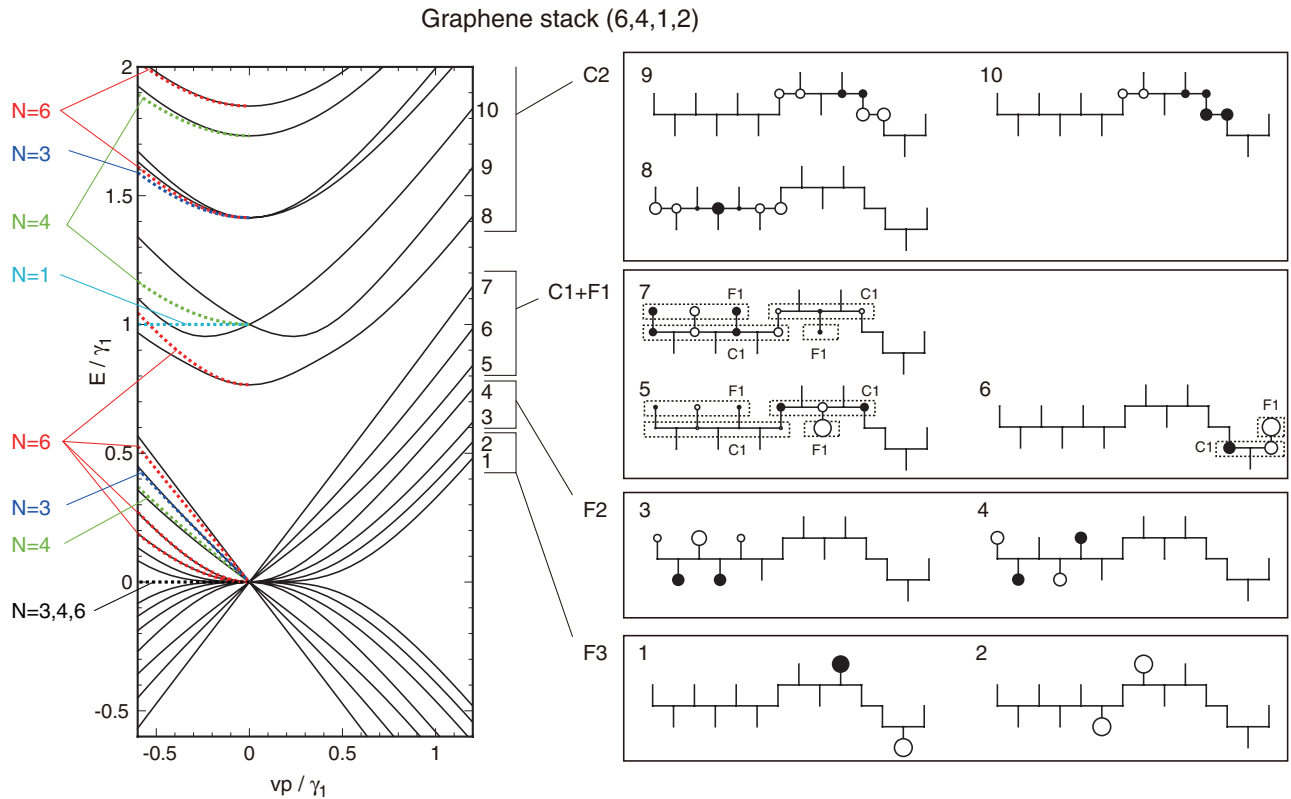}}
\caption{
Band structure and some wavefunctions
of the graphene stack $(6,4,1,2)$.
Dashed curves plotted for $p<0$ and $E>0$ represent
the energy bands of the decomposed incomplete graphites,
$N=6,4,1,$ and $2$ ($N=6$ and 2 are the end-section type).
}
\label{fig_band_6412}
\end{figure*}

The intersection coupling modifies the band mass
of C2 bands in Eq.\ (\ref{eq_band_C2}).
The strength of coupling is given by the product of $v p$
and the wave amplitudes at the end sites, Eq.\ (\ref{eq_endsite_amp}).
The mass change is irrelevant when
$N$ is large $(\gg 1)$ or $|q_l|$ is close to 0 or $\pi$.
When the energies of C2 bands in neighboring sections
happen to coincide at $p=0$, the two bands are strongly  hybridized
to give a linear $p$ term in addition to the original quadratic term.
This situation always occurs when the same $q_l$ is shared by the
neighboring sections.
In the example of Fig.\ \ref{fig_band_6412}, we see that
both of the sections $N=4$ and 1 have a C2 state of $q_l =\pi/3$,
and those two bands are strongly hybridized to
the 9th and 10th bands, which are linear at $p=0$.

\section{Statistical properties}

\subsection{Linear band velocity and quadratic band mass}

The velocity of linear bands
and the effective mass of quadratic bands
are important parameters characterizing
the low-energy spectrum which can be probed experimentally,
for example, by magneto-optical spectroscopy
\cite{Plochocka_et_al_2008a,Orlita_et_al_2009a}.
In Bernal-rhombohedral mixed graphene multilayer,
there are several frequently appearing values
of velocity and mass, each of which is associated with a
single incomplete graphite section or particular combinations of them.

\begin{figure*}
\centerline{\epsfxsize=0.95\hsize \epsffile{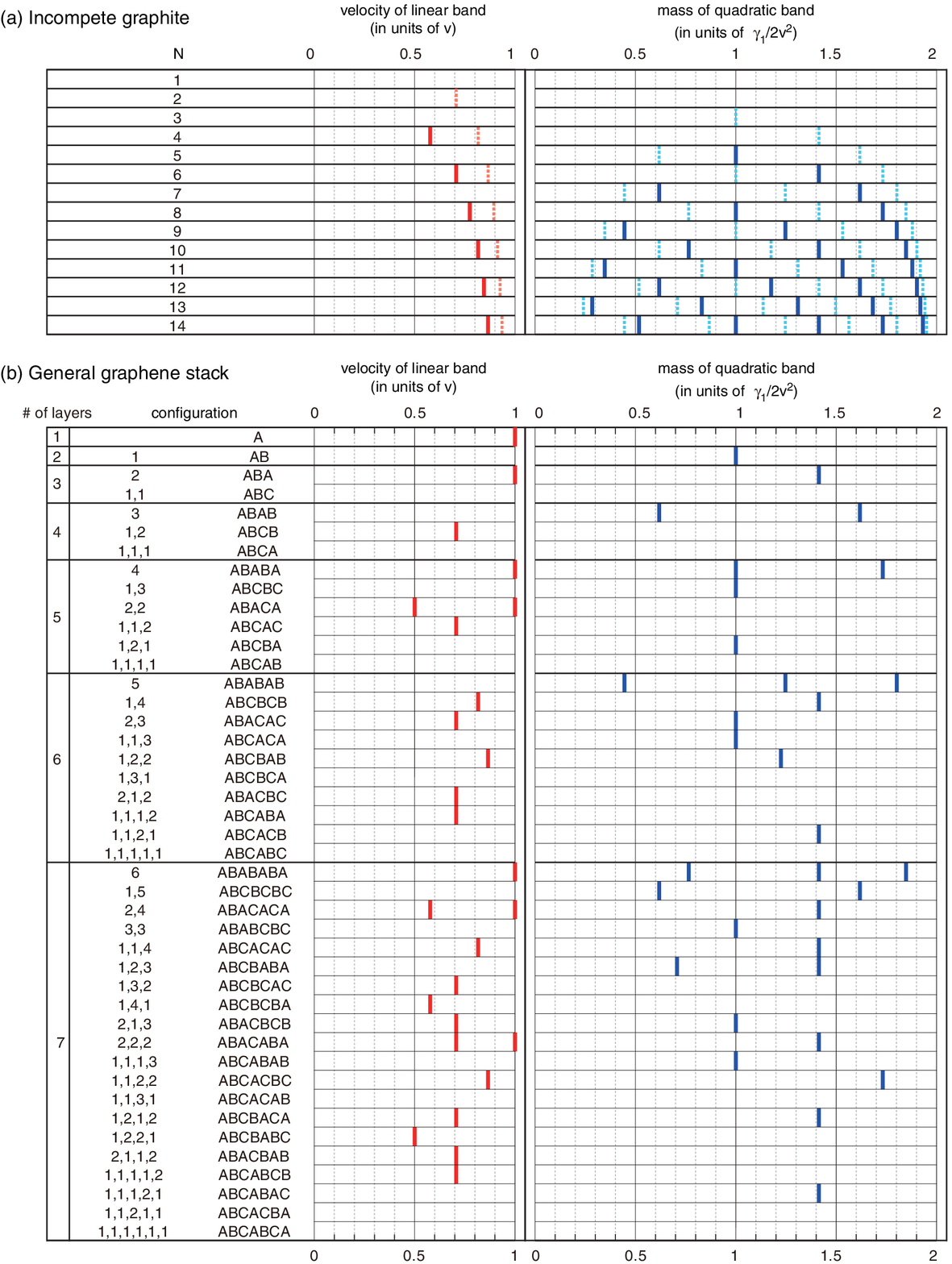}}
\caption{
(a) Lists of the linear band velocity and the quadratic band mass
in individual incomplete Bernal graphite sections from $N=1$ to $14$.
Solid and dashed bars represent the values of middle sections
and end sections, respectively.
(b) List of the same quantities for multilayer stacks
of all possible configurations from a single layer to 7 layers.
}
\label{fig_big_table}
\end{figure*}

For an isolated incomplete graphite,
the velocity of linear band (C1+F1) is given by
$\sqrt{(N-2)/(N+2)}v$ for a middle section
[Fig.\ \ref{fig_fragment}(b)],
and $\sqrt{N/(N+2)}v$
for an end section [Fig.\ \ref{fig_fragment}(c)].
The effective mass of the quadratic bands (F2)
is $|2(\gamma_1/v^2)\cos q'_l|$,
where $q_l'$ is given by Eqs.\ (\ref{eq_q_F}) and
(\ref{eq_q_F_end}) for a middle section
and an end section, respectively.
Fig.\ \ref{fig_big_table}(a) lists
the linear band velocity and the quadratic band mass
in incomplete graphite sections with some different $N$'s.
Solid and dashed bars represent the values of middle sections
and end sections, respectively.

Fig.\ \ref{fig_big_table}(b) presents the same quantities
of the mixed multilayer stacks
of all possible configurations from single layer to 7 layers.
There we derived the velocity and the mass from the
numerically calculated eigenenergies
of the original Hamiltonian Eq.\ (\ref{eq_H})
at small momenta. 
In multilayer $(2,3)$, for example,
we see that the linear band velocity is actually carried over from
an isolated graphite of $N=2$, and
the quadratic band mass is from $N=3$ (both of the end-section type)
which are listed in Fig.\ \ref{fig_big_table}(a).
For the case where even numbers appear consecutively
like $(2,4)$, on the other hand, the linear-band
velocities do not coincide with
those of the isolated systems as argued in the previous section.
The velocities of hybridized linear bands
are then calculated by diagonalizing Eq.\ (\ref{eq_C1+F1_mixed}).
Similarly, we see that the quadratic band mass
does not match with isolated ones,
when ``2'' appears somewhere in the middle, as $(1,2,3)$.
Then the hybridized mass is given by solving Eq.\ (\ref{eq_reconst1}).

The hybridized bands very often
have the same band dispersions as a single incomplete graphite section.
Tables \ref{tbl_v_and_m}(a) and \ref{tbl_v_and_m}(b)
list typical values of the linear band velocity
and the quadratic band mass, respectively,
with the corresponding configurations.
In the velocity table, the sequence ($\circ$,$a$,$b$,$\circ$) represents
the combination $a,b$ appearing in the middle of the multilayer,
and ($a$,$b$,$\circ$) is one located at the end.
The symbol $\circ$ represents an arbitrary sequence of numbers
in which no even number comes next to $a$ or $b$
(otherwise the linear bands are hybridized).
In the mass table, similarly,
 $\bullet$ represents any sequences
in which a number other than 2 comes next to $a$ and $b$.

\begin{table}
\begin{tabular}{|c|l|}
\hline
\parbox{23mm}{
\quad (a) velocity \\ (in units of $v$)
} &
\parbox{55mm}{configuration
} \\ \hline
 1/2 & ($\circ$,2,2,$\circ$)($\circ$,4,6,$\circ$)  \\ \hline
1/$\sqrt{3}$ & ($\circ$,4,$\circ$)($\circ$,4,2,4,$\circ$) \\ \hline
1/$\sqrt{2}$ & (2,$\circ$)($\circ$,6,$\circ$)($\circ$,2,4,$\circ$)($\circ$,2,2,2,$\circ$) \\ \hline
$\sqrt{3/5}$ & ($\circ$,8,$\circ$) \\ \hline
$\sqrt{2/3}$ & (4,$\circ$)($\circ$,10,$\circ$)($\circ$,4,6,$\circ$)($\circ$,2,4,2,$\circ$)($\circ$,4,2,4,$\circ$) \\ \hline
$\sqrt{3}/2$ & (6,$\circ$)(2,2,$\circ$)($\circ$,2,10,$\circ$)($\circ$,6,8,$\circ$)($\circ$,2,6,2,$\circ$) \\ \hline
\end{tabular}

\vspace{5mm}

\begin{tabular}{|c|l|}
\hline
\parbox{23mm}{
\quad (b) mass (in units
of $\gamma_1/2v^2$)
} &
\parbox{60mm}{configuration
} \\ \hline
 $(\pm1+\sqrt{5})/2$
&
\parbox{60mm}{
(5,$\bullet$)(10,$\bullet$)($\bullet$,7,$\bullet$)($\bullet$,2,2,7,$\bullet$)($\bullet$,5,2,5,$\bullet$)
}  \\ \hline
 $1$
& \parbox{60mm}{
 (3,$\bullet$)(6,$\bullet$)($\bullet$,5,$\bullet$)($\bullet$,8,$\bullet$)($\bullet$,3,2,3,$\bullet$)\\
($\bullet$,2,2,5,$\bullet$)($\bullet$,3,2,6,$\bullet$)($\bullet$2,2,8,$\bullet$)($\bullet$,5,2,5,$\bullet$)
}
  \\ \hline
$\sqrt{2}$
& \parbox{60mm}{
 (4,$\bullet$)(8,$\bullet$)(1,2,1$\bullet$)($\bullet$,6,$\bullet$)($\bullet$,10,$\bullet$)($\bullet$1,2,3,$\bullet$)\\
($\bullet$,1,2,7,$\bullet$)($\bullet$,2,2,6,$\bullet$)($\bullet$3,2,5,$\bullet$)($\bullet$,4,2,4,$\bullet$)
}
  \\ \hline
$\sqrt{3}$
& \parbox{60mm}{
 (6,$\bullet$)($\bullet$,8,$\bullet$)($\bullet$1,2,4,$\bullet$)($\bullet$,2,2,8,$\bullet$)($\bullet$,4,2,7,$\bullet$)\\
}
  \\ \hline
\end{tabular}
\caption{
(a) List of frequently-appearing velocity
of the low-energy linear band.
Symbol $\circ$ in ($\circ$,$a$,$b$,$\circ$)
represents an arbitrary sequence of numbers
in which no even number comes next to $a$ or $b$.
(b) List of frequently-appearing effective mass
of the low-energy quadratic band.
Symbol $\bullet$ in ($\bullet$,$a$,$b$,$\bullet$)
represents any sequences
in which a number other than 2 comes next to $a$ or $b$.
}
\label{tbl_v_and_m}
\end{table}

\subsection{Statistics of velocity and mass}

A realistic graphene multilayer system
is not always a pure system with a single stacking arrangement,
but is often composed of many small domains
with different stacking structures \cite{Ping_and_Fuhrer_2012}.
Nevertheless we expect to observe strong signals
from special band velocities or masses as argued above,
because they are statistically likely to appear in a random configuration.
Here we consider the distribution of band velocity and mass
ensemble-averaged over possible Bernal-rhombohedral mixed
configurations.
Apart from tight-binding parameters $\gamma_0$ and $\gamma_1$,
the only parameter to characterize the system is $p_s$,
the probability to have an $ABC$ stacking fault in every layer.
Namely, the system is a pure Bernal multilayer when $p_s=0$,
and pure rhombohedral multilayer when $p_s=1$.
Experimentally, the value of $p_s$ is not well known,
and it should depend on the multilayer growing process.
From the fact that rhombohedral structure is found by a few 10\%
out of natural graphite \cite{Lipson_and_Stokes_1942a},
we expect $p_s$ is typically a few times 0.1.

\begin{figure}
\centerline{\epsfxsize=0.88\hsize \epsffile{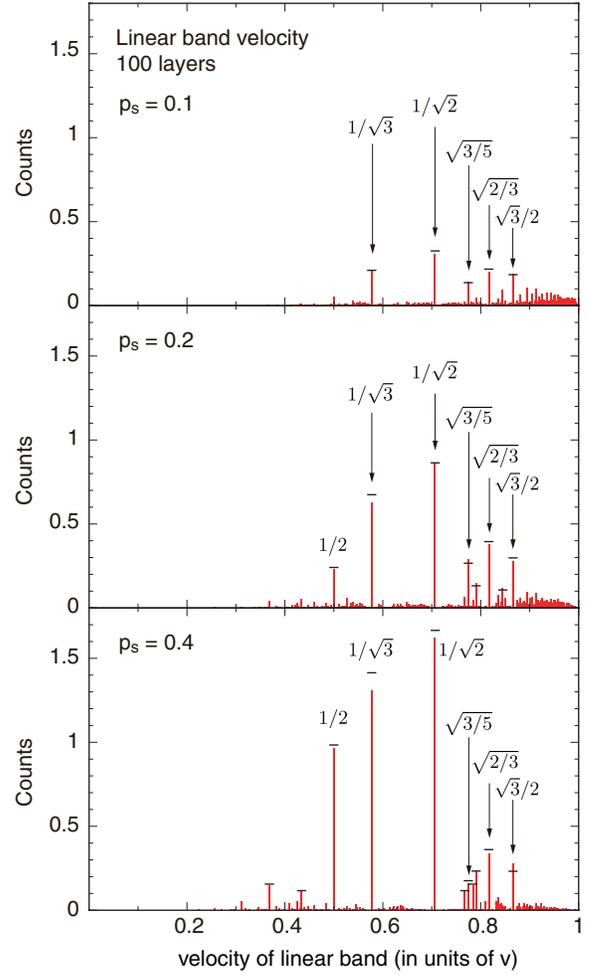}}
\caption{
Averaged distribution of the velocity of low-energy linear bands
in 100-layer Bernal-rhombohedral mixed graphite
with the stacking-fault probability $p_s=0.1$, 0.2, and 0.4.
Short horizontal bars attached to some major peaks
are obtained by the approximation (see the text).
}
\label{fig_v_stat}
\end{figure}

\begin{figure}
\centerline{\epsfxsize=0.85\hsize \epsffile{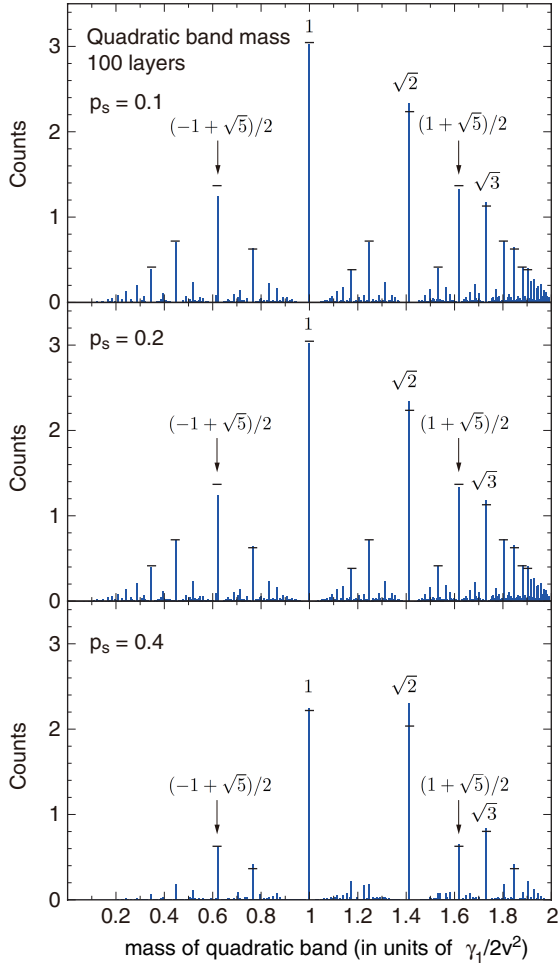}}
\caption{
Averaged distribution of effective mass of low-energy quadratic bands
similar to Fig.\ \ref{fig_v_stat}.
}
\label{fig_m_stat}
\end{figure}

Figs.\ \ref{fig_v_stat} and \ref{fig_m_stat} show
the distribution of the linear band velocity and
quadratic band mass in 100-layer systems
averaged over 1000 different configurations,
at $p_s=0.1$, 0.2, and 0.4.
The velocities and the masses are again obtained
from the eigenenergies of the original Hamiltonian Eq.\ (\ref{eq_H})
at small momenta.
The width of each histogram bin is $0.003v$ for velocity,
and $0.006(\gamma_1/2v^2)$ for mass.
The height represents the average
number of bands per a single 100-layer system.
We actually observe the strong peaks at the frequently-appearing values
listed in Table \ref{tbl_v_and_m}.
The relative heights
between different peaks depends significantly on $p_s$:
In particular, the velocity of $1/2$ (in units of $v$)
is almost absent in $p_s=0.1$,
but at $p_s=0.4$ it is more than half of
the tallest peak of $1/\sqrt{2}$.
It might be possible to estimate $p_s$ experimentally
for a given graphite sample, by comparing the
amplitudes of different peaks.

We can approximately estimate the peak heights using
simple probability calculation without taking the average
of the exact band structure.
Let us consider a large multilayer stack composed of
$N_{\rm tot}$ layers, in which the stacking fault takes place
with probability $p_s$.
The probability for a given section
(surrounded by stacking faults) to have length $N$
is written as
\begin{equation}
 P(N) = p_s(1-p_s)^{N-1}.
\end{equation}
The averaged length of a section is given by
\begin{equation}
 \bar{N} = \sum_{N=1}^{\infty} NP(N) = \frac{1}{p_s},
\end{equation}
and the average number of sections included in the whole system is
\begin{equation}
 \bar{M} = \frac{N_{\rm tot}}{\bar{N}} = p_s N_{\rm tot}.
\end{equation}

Now we consider the expected number of a certain sequence $(\circ,a,\circ)$
existing in the whole system. When we look at a particular section,
the probability that its length is $a$
and the left- and right-neighboring sections are both odd numbers
at the same time, is given by $P_{\rm odd}P(a)P_{\rm odd}$,
where $P_{\rm odd} = P(1)+P(3)+P(5)+\cdots = 1/(2-p_s)$.
The number of the sequence $(\circ,a,\circ)$ in the system
is then obtained by multiplying this with the number of sections,
$\bar{M}$, namely,
\begin{equation}
 n(\circ,a,\circ) \approx p_s N_{\rm tot}  P(a) P_{\rm odd}^2.
\label{eq_n_middle}
\end{equation}
For the end section $(a,\circ)$, similarly, we have
\begin{equation}
 n(a,\circ) \approx 2 P(a) P_{\rm odd},
\label{eq_n_end}
\end{equation}
where the factor 2 comes from two ends of the whole system.
For the combinate sequence $(\circ,a,b,\circ)$,
$P(a)$ is just replaced with $P(a)P(b)$ in above equations.
For the case $(\bullet,a,\bullet)$ appearing
in the effective mass table,
$P_{\rm odd}$ is replaced with $1-P(2)$.

The population of a given velocity/mass can be estimated by
summing up these numbers over all the corresponding sequences.
For example, the number of bands with velocity $1/\sqrt{3}$
is approximated by $n(\circ,4,\circ) + n(\circ,4,2,4,\circ)$,
according to Table \ref{tbl_v_and_m}(a).
We calculate the populations of several representative velocities/masses
in this manner, and show the estimated values as short horizontal bars
in Figs.\ \ref{fig_v_stat} and \ref{fig_m_stat}.
While we took only a finite number of the configurations
giving relatively large contributions,
the approximation actually works quite well.
We note that the peak diagrams like
 Figs.\ \ref{fig_v_stat} and \ref{fig_m_stat}
are not universal but depend on the total number of layers $N_{\rm tot}$,
because
the contribution of the middle section is proportional to
$N_{\rm tot}$ as in Eq.\ (\ref{eq_n_middle}), while that of
the end section, Eq.\ (\ref{eq_n_end}), is not.
In very large multilayer stacks with $N_{\rm tot} >\sim 1000$,
the end section component becomes negligible,
and all the peaks just scale in proportional to $N_{\rm tot}$.

\subsection{Optical absorption spectrum}

Frequently appearing band structures
give rise to striking signals in the optical absorption spectrum.
The optical absorption is related to the dynamical conductivity,
which is given by within the linear response,
\begin{eqnarray}
 \sigma(\omega) =
\frac{e^2\hbar}{iS}
\sum_{\alpha,\beta}
\frac{f(\vare_\alpha)-f(\vare_\beta)}{\vare_\alpha- \vare_\beta}
\frac{| \langle\alpha | v_x | \beta\rangle |^2}
{\vare_\alpha- \vare_\beta+\hbar\omega+i\delta},
\label{eq_sigma_xx}
\end{eqnarray}
where $S$ is the area of the system,
$v_x = \partial {\cal H}/\partial p_x$ is the velocity operator,
$\delta$ is the positive infinitesimal,
$f(\vare)$ is the Fermi distribution function,
and $| \alpha\rangle$ and $\vare_\alpha$ describe the eigenstate and
the eigenenergy of the system.
The transmission of light incident perpendicular to a two-dimensional
system is given by \cite{Ando_1975a}
\begin{equation}
 T = \Big| 1+ \frac{2\pi}{c}\sigma(\omega) \Big|^{-2}.
\label{eq_Transmission}
\end{equation}

\begin{figure}
\centerline{\epsfxsize=1.\hsize \epsffile{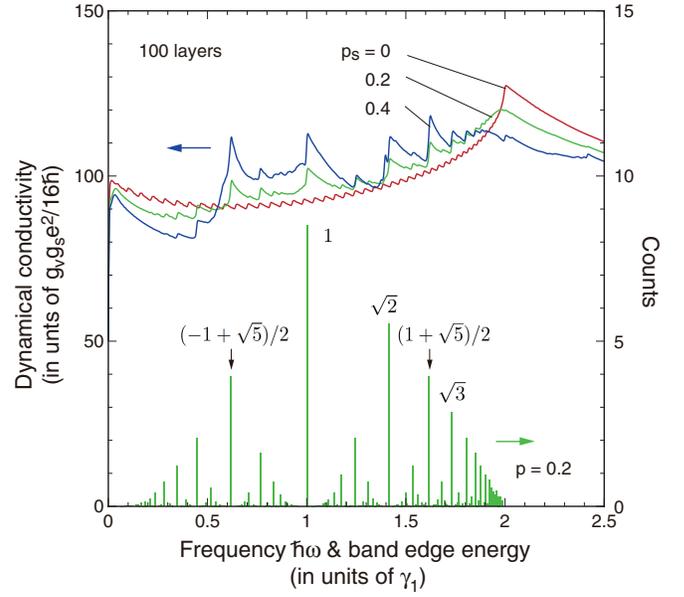}}
\caption{
Averaged dynamical conductivity
of 100-layer Bernal-rhombohedral mixed graphites
with stacking-fault probability $p_s=0,0.2$ and 0.4.
Histogram at the bottom is
the averaged distribution of the band edges
(band energies at zero momentum) in $p_s=0.2$.
}
\label{fig_opt}
\end{figure}

Here we calculate the dynamical conductivity
of 100-layer graphites averaged over 1000
different configurations at fixed
stacking-fault probability $p_s$.
We set the Fermi energy at the charge neutral point, $\vare_F=0$.
Fig.\ \ref{fig_opt} shows the dynamical conductivity
as a function of frequency $\omega$ calculated for $p_s=0,$ 0.2 and 0.4.
The peaks characterizing the spectrum mainly come
from the transition between $\vare=0$ and the band edge of the C1 band,
$\vare = 2\gamma_1 \cos q_l$.
The system at $p_s=0$ is a pure Bernal-stacked 100-layer graphite,
and the absorption edge appears at
$\hbar \omega = 2\gamma_1\cos q_l$ with
$q_l = l\pi/101$ $(l=1,2,\cdots,100)$.
Indeed, we see a regular series of small peaks for
$\hbar \omega < 2\gamma_1$.
A major peak at $\hbar\omega = \gamma_1$
comes from  densely distributed absorption edges at $q_l\sim 0$.

For finite $p_s$, on the other hand,
the spectrum is rather regarded as
a summation over isolated incomplete graphites of finite layers,
as long as the total number of layers is much larger than
the averaged length of the Bernal sections.
The histogram in Fig.\ \ref{fig_opt} shows
the distribution of the energy of
C1 band edge $\vare = 2\gamma_1 \cos q_l$
in a 100-layer system at $p_s=0.2$.
We clearly see that the major peaks appearing in the dynamical
conductivity correspond to the absorption edges of
frequently appearing band structures.

\section{Conclusion}

We studied the electronic structures of
general Bernal-rhombohedral mixed graphene multilayers
with arbitrary configurations.
We showed that every low-energy eigenstate is localized
in a finite Bernal section bound by rhombohedral-type stacking faults,
and the spectrum is well approximated by a collection of
the spectra of independent sections, categorized
into linear, quadratic and cubic (or higher order) bands.
We found that the ensemble-averaged spectrum is not smooth,
but exhibits a number of discrete structures
originating from finite Bernal sections or their combinations
which are statistically likely to appear in a random configuration.
In the low-energy region, in particular,
there are frequently-appearing linear bands and quadratic bands
with specific band velocities or curvatures.
In the higher energy region away from the Dirac point,
band edges are likely to appear at some particular energies.
Those discrete properties may be detected
in general graphite samples using experimental techniques such as
optical or magneto-optical absorption spectroscopy,
and those observations would be
useful to probe the stacking structure of graphene multilayers.

\section{Acknowledgments}

This project has been funded by JST-EPSRC Japan-UK Cooperative Programme Grant EP/H025804/1.

\appendix

\section{Eigen energies of F2 and F3 states}
\label{sec_app1}

Here we show that F2 and F3 states defined in the text
are approximate eigen states
with eigen energies Eqs.\ (\ref{eq_band_F2}) and
(\ref{eq_band_F3}), respectively,
in an independent incomplete Bernal graphite.
Let us consider a $N$-layer middle section
shown in Fig.\ \ref{fig_fragment}(b),
and assign the site indices $a_j (j=0,1,\cdots,N)$
and $b_j (j=1,\cdots,N-1)$ as in the figure.
The Schr\"{o}dinger equation on the layer $j=1,\cdots,N$
is written as
\begin{eqnarray}
\!\!\!\!\!\!\!\!\!&&\vare f(b_j) = v p e^{i\theta_j} f (a_j)
\label{eq_app_sch1}\\
\!\!\!\!\!\!\!\!\!&&  \vare f(a_j) = v p e^{-i\theta_j} f (b_j)
+ \gamma_1 [f(a_{j-1}) + f(a_{j+1})],
\label{eq_app_sch2}
\end{eqnarray}
and on $j=0$ and $N$ as
\begin{eqnarray}
&&  \vare f(a_0) = \gamma_1 f(a_1),
\nonumber\\
&&  \vare f(a_N) = \gamma_1 f(a_{N-1}),
\label{eq_app_sch3}
\end{eqnarray}
where $\vare$ is the eigen energy.

For the F2 state, the exact wave function
is given by Eq.\ (\ref{eq_wave_F}),
plus a correction due to the first order perturbation in the momentum $p$.
The wave amplitudes on the free sites are of 0th order
and given as in Eq.\ (\ref{eq_wave_F}),
\begin{eqnarray}
&& f(b_j) = C e^{i\theta_j} \sin q'_l (j-1),
\end{eqnarray}
where $C$ is a constant. The amplitudes on the chained sites are
then derived from Eq.\ (\ref{eq_app_sch1}) as
\begin{eqnarray}
&& f(a_j) =  \frac{\vare}{vp} C \sin q'_l(j-1).
\end{eqnarray}
For $j=1,\cdots,N-1$,
we substitute above $f(a_j)$ for Eq.\ (\ref{eq_app_sch2}) to find
\begin{eqnarray}
 \vare^2 = (2\gamma_1\cos q'_l)\vare + v^2p^2,
\end{eqnarray}
giving
\begin{eqnarray}
 \vare \approx -\frac{v^2p^2}{2\gamma_1\cos q'_l}
\end{eqnarray}
as long as $vp \ll 2\gamma_1\cos q'_l$.
For the end sites $j=0$ and $j=N$, on the other hand,
Eq.\ (\ref{eq_app_sch3}) yields to
\begin{eqnarray}
 \frac{\vare^2}{vp} \sin q_l' = 0,
\end{eqnarray}
which stands within $O(p^2)$
since $\vare \propto p^2$.

For F3 states, it is straightforward to show that
the exact eigenstates at finite $p$ are explicitly written as
\begin{eqnarray}
&& f(b_1) = 1, \quad f(a_0) = -\frac{vp e^{-i\theta_1}}{\gamma_1};
\quad {\rm for (F3,L)}
\nonumber \\
&& f(b_{N-1}) = 1, \quad f(a_N) = -\frac{vp e^{-i\theta_{n-1}}}{\gamma_1};
\quad {\rm for (F3,R)}
\nonumber\\
\end{eqnarray}
and the eigen energies are exactly zero.

\section{Inter-section coupling by $N_i=2$}

\label{sec_app2}

We consider the hybridization of F2 and F3 states
in a series of incomplete Bernal sections $(N_{i-1},2,N_{i+1})$,
as illustrated in Fig.\ \ref{fig_connect}(b),
to derive the equation (\ref{eq_reconst1})
for the reconstructed wavenumber $q_l'$.
We assign the site indices $(a_i,b_i)$, $(a'_i,b'_i)$
and $(a''_i,b''_i)$ for the section $N_{i-1}$, 2, and $N_{i+1}$,
respectively, as shown in Fig.\ \ref{fig_connect}(b).
We assume either $N_{i\pm2}$ (out of the figure) is not 2, 
and then we can neglect the coupling with them.

We consider a small momentum $p \ll \gamma_1/v$,
and treat it as a perturbation.
The 0th order wavefunction for a hybridized state is expressed as
a linear combination of F2 and F3 states of the sections $N_{i-1}$ and $N_{i+1}$
having a node at $b_1$ and $b''_{N_{i+1}-1}$, respectively,
and C1 state in the middle.
Then the wave function is written as
\begin{eqnarray}
&& f(b_j) = C e^{i\theta_j}\sin q(j-1)
\nonumber\\
&& f(b''_j) = C'' e^{i\theta''_j}\sin q(j-N_{i-1}+1)
\nonumber\\
&& f(a'_0) = -f(a'_2) = C',
\label{eq_app_f1}
\end{eqnarray}
where $C,C',C''$ and $q$ are quantities to be solved,
and $\theta_j (\theta''_j)$ is
defined by $e^{i\theta_j}(e^{i\theta''_j}) = (p_x\pm i p_y)/p$
when $b_j(b''_j)$ is $B$-site and $A$-site, respectively.

The amplitudes at chained sites $a_j$ and $a''_j$, which are linear in
$p$,
can be obtained in the exactly same way as in Appendix \ref{sec_app1} as
\begin{eqnarray}
&& f(a_j) =  \frac{\vare}{vp} C \sin q(j-1)
\nonumber\\
&& f(a''_j) = \frac{\vare}{vp} C''\sin q(j-N_{i-1}+1),
\label{eq_app_f2}
\end{eqnarray}
where $\vare$ is the eigen energy.
The Schr\"{o}dinger equations at $a_{N_{i-1}}$ and $a''_{0}$
are given by
\begin{eqnarray}
 && v p e^{-i\theta_{N_{i-1}}} f (a'_0)
+ \gamma_1 f (a_{N_{i-1}-1})  = \vare f(a_{N_{i-1}}), \nonumber\\
 && v p e^{-i\theta''_{0}} f (a'_2)
+ \gamma_1 f (a''_{1})  = \vare f(a''_{0}).
\end{eqnarray}
By substituting the wave amplitudes, we have
\begin{eqnarray}
 && C' = \frac{\vare}{v^2p^2} C
\left[
\vare\sin q(N_{i-1}-1) - \gamma_1\sin q(N_{i-1}-2)
\right] \nonumber\\
&&  C' = -\frac{\vare}{v^2p^2} C''
\left[
\vare\sin q(N_{i+1}-1) - \gamma_1\sin q(N_{i+1}-2)
\right]
\nonumber\\
\label{eq_app_1}
\end{eqnarray}
Also, the Schr\"{o}dinger equations at $a'_{0}$ and $a'_{2}$,
\begin{eqnarray}
 && v p e^{i\theta_{N_{i-1}}} f(a_{N_{i-1}}) + \gamma_1 f(a'_1)
  = \vare f(a'_0) \nonumber\\
 && v p e^{i\theta''_{0}} f(a''_{0}) + \gamma_1 f(a'_1)
  = \vare f(a'_2),
\end{eqnarray}
leads to
\begin{eqnarray}
C' = \frac{1}{\sqrt{2}}\left[
C \sin q(N_{i-1}-1) + C''\sin q(N_{i+1}-1)
\right].
\label{eq_app_2}
\end{eqnarray}
The wavenumber $q$ and the relative wave amplitudes $C'/C$ and $C''/C$
are obtained by solving Eqs.\ (\ref{eq_app_1}) and (\ref{eq_app_2}).
After some algebra, we obtain the equation for $q = q'_l$
as Eq.\ (\ref{eq_reconst1}).
While the argument above is only valid when $N_{i\pm 1} \geq 3$,
we can show that Eq.\ (\ref{eq_reconst1}) stand also when
either or both $N_{i\pm 1}$ is 1 or 2.

\end{document}